%% file: main.tex
\documentclass[sigconf]{acmart}

\usepackage{amsmath}
\usepackage{color}
\usepackage{subfigure}
\usepackage{array}
\usepackage{{inputenc}}
\usepackage{balance}
\usepackage[ruled,vlined]{algorithm2e}
\usepackage{algpseudocode}
\usepackage{multirow,tabularx}
\usepackage{booktabs,longtable}
\usepackage{graphicx}
\usepackage{bbm}
\usepackage{enumitem}
\usepackage{verbatim}
\usepackage{amsfonts}
\usepackage{hyperref}
\hypersetup{colorlinks=false,linkcolor=blue,urlcolor=blue,citecolor=red}

\newcommand{\Lapl}{\mathbf{\mathop{\mathcal{L}}}}

\newcommand{\Mat}[1]{\textbf{#1}}

\newcommand{\Space}[1]{\mathbb{#1}}
\newcommand{\Set}[1]{\mathcal{#1}}

\newcommand{\ie}{\emph{i.e., }}
\newcommand{\eg}{\emph{e.g., }}

\newcommand{\wrt}{\emph{w.r.t. }}

\newcommand{\argtop}[1]{\mathop{\mathrm{argTop{#1}}}}

\AtBeginDocument{%
  \providecommand\BibTeX{{%
    \normalfont B\kern-0.5em{\scshape i\kern-0.25em b}\kern-0.8em\TeX}}}

\copyrightyear{2023}
\acmYear{2023}
\setcopyright{acmlicensed}\acmConference[WWW '23]{Proceedings of the ACM
Web Conference 2023}{May 1--5, 2023}{Austin, TX, USA}
\acmBooktitle{Proceedings of the ACM Web Conference 2023 (WWW '23), May
1--5, 2023, Austin, TX, USA}
\acmPrice{15.00}
\acmDOI{10.1145/3543507.3583303}
\acmISBN{978-1-4503-9416-1/23/04}

\begin{document}

\title{Towards Explainable Collaborative Filtering with Taste Clusters Learning }

\author{Yuntao Du}
\affiliation{%
    \institution{Zhejiang University}
    \city{Hangzhou}
    \country{China}
}
\email{ytdu@zju.edu.cn}

\author{Jianxun Lian}
\affiliation{%
    \institution{Microsoft Research Asia}
    \city{Beijing}
    \country{China}
}
\email{jianxun.lian@outlook.com}

\author{Jing Yao\\Xiting Wang}
\affiliation{%
    \institution{Microsoft Research Asia}
    \city{Beijing}
    \country{China}
}
\email{{jingyao, xitwan}@microsoft.com}


\author{Mingqi Wu}
\affiliation{%
    \institution{Microsoft Gaming}
    \city{Redmond}
    \country{United States}
}
\email{mingqi.wu@microsoft.com}

\author{Lu Chen \\ Yunjun Gao}
\affiliation{%
    \institution{Zhejiang University}
    \city{Hangzhou}
    \country{China}
}
\email{{chenlu, gaoyj}@zju.edu.cn}


\author{Xing Xie}
\affiliation{%
    \institution{Microsoft Research Asia}
    \city{Beijing}
    \country{China}
}
\email{xingx@microsoft.com}
\renewcommand{\shortauthors}{Du et al.}

\begin{abstract}
Collaborative Filtering (CF) is a widely used and effective technique for recommender systems. In recent decades, there have been significant advancements in latent embedding-based CF methods for improved accuracy, such as matrix factorization, neural collaborative filtering, and LightGCN. However, the explainability of these models has not been fully explored. Adding explainability to recommendation models can not only increase trust in the decision-making process, but also have multiple benefits such as providing persuasive explanations for item recommendations, creating explicit profiles for users and items, and assisting item producers in design improvements.

In this paper, we propose a neat and effective Explainable Collaborative Filtering (ECF) model that leverages interpretable cluster learning to achieve the two most demanding objectives: (1) Precise - the model should not compromise accuracy in the pursuit of explainability; and (2) Self-explainable - the model's explanations should truly reflect its decision-making process, not generated from post-hoc methods.
The core of ECF is mining taste clusters from user-item interactions and item profiles. We map each user and item to a sparse set of taste clusters, and taste clusters are distinguished by a few representative tags. The user-item preference, users/items' cluster affiliations, and the generation of taste clusters are jointly optimized in an end-to-end manner. Additionally, we introduce a forest mechanism to ensure the model's accuracy, explainability, and diversity.  To comprehensively evaluate the explainability quality of taste clusters, we design several quantitative metrics, including in-cluster item coverage, tag utilization, silhouette, and informativeness. Our model's effectiveness is demonstrated through extensive experiments on three real-world datasets.
\end{abstract}




\maketitle

\input{1_intro}

\input{2_methodology}

\input{3_exp}

\input{4_related}

\input{5_conclusion}

\begin{acks}
This work was supported in part by the NSFC under Grants No. (62025206, 61972338, and 62102351), and the Ningbo Science and Technology Special Innovation Projects with Grant Nos. 2022Z095 and 2021Z019.
Yunjun Gao is the corresponding author of the work.
\end{acks}

\bibliographystyle{ACM-Reference-Format}
\balance

\bibliography{ref}

\input{6_appendix}

\end{document}

%% file: 1_intro.tex
\section{Introduction}
Collaborative filtering (CF) is an effective and widely used technique for recommender systems. The rationale behind CF is that the user's potential interests can be inferred from a group of like-minded users' behaviors (\eg user-based CF), or, a set of similar items to the user's behavior history (\eg item-based CF). 
Existing methods for CF can be roughly categorized into two types: memory-based CF and model-based CF. Memory-based CF methods~\cite{www01itemcf} explicitly find the neighborhood of users/items for preference inference.  These methods, although concise and effective, usually suffer from memory consumption, (inference) computational cost, and data sparsity issues. In contrast, model-based CF uses machine learning methods to help model the relationship between users and items. In the past decades, the research trends have been mainly focusing on model-based CF,  with methods evolving from latent factorization-based methods (such as MF~\cite{09mf} and SVD++~\cite{kdd08svd}) to neural embedding-based methods (such as NCF~\cite{www17ncf} and CDAE~\cite{wsdm16cdae}), and to recently neural graph-based methods (such as LightGCN~\cite{sigir20lighGCN} and UltraGCN~\cite{cikm21ultragcn}). The foundation model in this direction is to represent users and items with high-quality latent factor embeddings, so that users' preference towards items can be decoded from the latent embeddings.

A notable drawback of latent factor-based CF is the lack of transparency and explainability. Knowing the decision logic of recommender algorithms rather than using them as black boxes is important on multiple aspects, such as assisting developers in model debugging and abnormal case studies or generating persuasive explanations to promote the recommended items and increase conversion rates. Many prior attempts have been made to achieve this goal. To name a few representative works, \cite{recsys13hft} combines a latent matrix factorization model with a latent topic model, then each dimension of user/item embeddings can be associated with a textual topic mined from users' reviews. \cite{sigir14efm} proposes the Explicit Factor Model (EFM), which extracts aspect features from textual reviews and makes predictions by combining scores from a latent factor model and from aspect-level interest matching. \cite{ijcai20amcf} proposes a feature mapping paradigm. By mapping latent embeddings to interpretable aspect features, an uninterpretable model can now provide explainability. However, we argue that explainable CF is still an open question, because existing solutions fail to satisfy at least one of the properties: (P1,flexibility) the dimension of latent embeddings and the number of interpretable features/topics do not necessarily match (\eg \cite{recsys13hft} fails on this); (P2,coherence) a model's interpretable modules and predictive modules should be harmoniously integrated for decision making, instead of functioning as isolated modules, which may lead to divergent outcomes (\eg \cite{sigir14efm} fails on this);    (P3, intrinsic explainability) a model is inherently understandable and interpretable, instead of relying on a post-hoc model for explanations (\eg \cite{ijcai20amcf} fails on this). 

In this paper, we propose a neat yet effective Explainable Collaborative Filtering (ECF) framework with the goal of accurate and explainable recommendations, while satisfying all three properties. At the core of ECF is mining various taste clusters. We assume that items have some explicit features such as tags or aspects (hereinafter we will use tags for unification). A taste cluster is a group of items which are not only similar in users' latent interest space, but also explicitly share some common tags, so that the overlapping tags can be selected as descriptive interpretation for the taste cluster. Both users and items are mapped to multiple taste clusters. On the one hand, item recommendations can be made by measuring the coherence between users' and items' taste cluster affiliations; on the other hand, the descriptive tags interpret what determines users' preference towards items. We also design a new set of quantitative metrics to evaluate the quality of ECF beyond the accuracy, covering:  
(1) In-cluster item coverage; (2) Tag utilization; (3) Silhouette; (4) Tag informativeness, with the goal of discovering the discriminability  of taste clusters as well as the exactitude, diversity, and uniqueness of clusters interpretability in a holistic manner. 
The ECF framework has many more application potentials beyond the user-item prediction and interpretation. For example, the affiliation of users to taste clusters can be used for user profiling, empowering web-scale services like ads audience targeting and audience extension. The affiliation of items to taste clusters can help missing tags discover for items, automatic item topic generation and theme-wise item recommendations (such as the playlist recommendation in online music service). 

However, the implementation and optimization of ECF is a non-trivial task. Main challenges include: (1) how to construct an end-to-end model, so that taste cluster generation, tags selection, user- and item-cluster affiliations can be optimized jointly; (2) how to achieve good performance on both accuracy and explainability simultaneously, instead of sacrificing one for the other; (3) how to guarantee the diversity of generated taste clusters. To this end, we design an embedding-based model to generate taste clusters with discriminative tags and establish the sparse mapping between users/items and clusters. We further devise the forest mechanism for ECF, which can not only substantially lift the accuracy, but also provides a diverse set of taste clusters. For each targeting merit (such as exactitude, diversity, and uniqueness of clusters interpretability) we design a corresponding objective function, so that the whole model can be trained in an end-to-end manner.


In summary, our main contributions include: 
\begin{itemize}[leftmargin=*]
    \item We present a neat yet effective explainable collaborative filtering framework, called ECF
    ~\footnote{The codes are available at \url{https://github.com/zealscott/ECF}.}, which leverages interpretable taste clusters and sparse user- and item-cluster affiliations for recommendation in a flexible, coherent, and self-explainable way.
    \item We propose an optimization method for ECF to learn high quality taste clusters with informative tags and sparse affiliations simultaneously in an end-to-end manner.
    \item We design a new set of quantitative metrics to evaluate the explainability quality of taste clusters comprehensively.
    \item We conduct extensive experiments on three real-world datasets with various state-of-the-art methods to demonstrate the effectiveness of ECF.
\end{itemize}

%% file: 2_methodology.tex
\section{Methodologies}
In this section, we first outlines the task definition for explainable recommendations and introduces the framework of Explainable Collaborative Filtering (ECF). We then delve into the methodology of the end-to-end optimization process for training ECF and establish a comprehensive set of metrics to quantitatively evaluate the effectiveness of explainability. 

\subsection{Problem Formulation}
 
\vspace{3pt}
\noindent
\textbf{User-item Interactions and Item Tags.} We focus on the implicit feedback~\cite{uai09BPRloss} in recommendation, where a user's preferences are inferred through implicit signals such as clicks or play actions. Let $\Set{U}$ be a set of users and $\Set{I}$ a set of items. Let $\Mat{Y}\in \Space{R}^{|\Set{U}|\times |\Set{I}|}$ as the user-item interaction matrix, where $y_{ui} = 1$ means that there is an interaction between user $u$ and item $i$, otherwise $y_{ui} = 0$. Besides, we assume there is a set of tags $\Set{T}$ for items, and each item $i$ has a subset of tags $\Set{T}_i \subset \Set{T}$ to describe its genre and features.

\vspace{3pt}
\noindent
\textbf{Task Description.} Given the interactions data $\Mat{Y}$ and items' tags $\Set{T}$, the task of explainable recommendation is to (1) predict how likely user $u$ adopts item $i$, and (2) explain what determines the recommendation of item $i$ to user $u$ with the descriptions from $\Set{T}$.

\subsection{The Proposed ECF framework} \label{sec:framework}
We propose an ECF framework for explainable recommendations with collaborative filtering.
The core of ECF is mining various taste clusters which represent specific interest patterns. We first give the definition of taste clusters and show how to leverage them for recommendation and explanation. Then we detail the proposed method to generate sparse affiliations between users/items and taste clusters. The pseudocode of ECF is shown in the Appendix~\ref{appendix:pseudocode}.

A \textbf{Taste Cluster} is a group of items $c = \{i_1, i_2,...,i_c\}$ that uncovers a certain type of underlying interest for users and can be easily identified through shared tags among the items within the cluster. So, each taste cluster is associated with a set of tags $\Set{T}_c = \{t_{1}, t_{2},...,t_{k}\}$, which are derived from corresponding item sets and acts as the descriptive interpretation for the cluster.

Assuming there is a set of taste clusters $\Set{C} = \{c\}$, we map users/items to these taste clusters with \textit{sparse affiliation matrix}. To be specific, let $\Mat{A}\in\Space{R}^{|\Set{U}|\times |\Set{C}|}$ and $\Mat{X}\in\Space{R}^{|\Set{I}|\times |\Set{C}|}$ be the affiliation matrices between users/items and taste clusters, respectively. Each entry $a_{uc}$ in $\Mat{A}$ denotes the preference degree of user $u$ to taste cluster $c$, and each entry $x_{ic}$ denotes the relatedness between item $i$ and taste cluster $c$. 
To enhance the clarity and persuasiveness of the explanation, the affiliation matrix should be kept sparse, meaning only the most pertinent entries have non-zero values, making it possible to clearly identify a user's preferences and item affiliations using only a few taste clusters.

After that, both users and items are mapped to multiple taste clusters, and ECF can make user-item recommendation and explanation with taste clusters and affiliations as follows.

\vspace{3pt}
\noindent
\textbf{Item recommendation.}
We assume that a user’s decision about whether to make a purchase/click is based on her preference with taste clusters, as well as the item’s relatedness with them. Thus, the prediction score of user $u$ and item $i$ can be calculated by multiplying their affiliations:
{\setlength{\abovedisplayskip}{3pt}
\setlength{\belowdisplayskip}{3pt}
\begin{gather}\label{equ:ecf-prediction}
\hat{y}_{ui} = \mathrm{sparse\_dot} (\Mat{a}_u, \Mat{x}_i),
\end{gather}}where $\mathrm{sparse\_dot}(\cdot)$ denotes sparse dot product for 
$\Mat{a}_u$ and $\Mat{x}_i$.

\vspace{3pt}
\noindent
\textbf{Personalized explanation.}
For each prediction $\hat{y}_{ui}$, ECF is able to generate explanation by measuring the coherence between users' and items' taste cluster affiliations. Specifically, the overlapped taste clusters are first derived from affiliation matrix:
{\setlength{\abovedisplayskip}{3pt}
\setlength{\belowdisplayskip}{3pt}
\begin{gather}\label{equ:ecf-explanation-overlap}
\Set{C}_{ui} = S(\Mat{a}_u) \cap S(\Mat{x}_i) ,
\end{gather}}where $S(\cdot)$ is an index selector to collect corresponding clusters where the affiliation between users/items and taste clusters exists, and $\Set{C}_{ui}$ denotes the set of overlapped taste clusters for user $u$ and item $i$. Thus, the descriptive tags of taste clusters in $\Set{C}_{ui}$ are used to interpret what determines user $u$ preference toward item $i$. 
Moreover, importance score $w_{ui}^c$ is introduced to quantify the contribution of each taste cluster in $\Set{C}_{ui}$:
{\setlength{\abovedisplayskip}{3pt}
\setlength{\belowdisplayskip}{3pt}
\begin{gather}\label{equ:ecf-explanation-weight}
w_{ui}^c = a_{uc}\times x_{ic} .
\end{gather}}Therefore, ECF is able to explain the prediction decision by using the descriptive tags of overlapped taste clusters $\Set{C}_{ui}$ and their corresponding importance scores $w_{ui}^c$. Due to the sparsity of affiliation matrix, the size of overlapped taste clusters is small, so that we could easily derive the readable decision logic of ECF by investigating only a few taste clusters.
And the final explanation results can be constructed by existing methods (\eg template-based explanation~\cite{sigir14efm}) or explanation paths from user to taste clusters and items (demonstrated in Section~\ref{exp-case-study-rec}).

Despite the simplicity, it is still unclear how to obtain the desired taste clusters and sparse affiliation matrix. This is a non-trivial task since performing clustering algorithm (\eg K-means) for items or blindly collecting items with same tags would fail to encode both user interest and item profiles simultaneously (see Section~\ref{sec:eval-explanation} for comparison). Besides, directly learning the affiliation matrix from data is unable to train due to its sparsity nature. 
Instead, we represent items and taste clusters as embeddings, and measure their affiliations with cosine similarity:
{\setlength{\abovedisplayskip}{3pt}
\setlength{\belowdisplayskip}{3pt}
\begin{gather}\label{equ:item-cluster-sim}
    \tilde{x}_{ic} = \cos (\Mat{v}_i, \Mat{h}_c),
\end{gather}}where $\Mat{v} \in \Space{R}^{d}$ and $\Mat{h} \in \Space{R}^{d}$ are the embeddings of items and clusters, and $d$ is the embedding size. To clearly identify the belongings between items and clusters, we only consider the Top-$m$ clusters for each item:
{\setlength{\abovedisplayskip}{3pt}
\setlength{\belowdisplayskip}{3pt}
\begin{gather}
    \label{equ:mask-item-cluster-sim}
    m_{ic} = \left\{\begin{array}{cl}
    1 & \text { if } c\in \argtop{m} (\tilde{\Mat{x}}_i) \\
    0 & \text { otherwise }
    \end{array}\right. \\
    \Mat{x}_i = \sigma (\tilde{\Mat{x}}_i ) \odot \Mat{m}_i , \label{equ:item-cluster-sparse}
\end{gather}}where $\odot$ denotes Hadamard product, $\sigma(\cdot)$ is the sigmoid function. In the above equations, we first make Top-$m$ selection for each item according to the similarity scores, and apply the sigmoid function to indicate the relatedness of items and clusters with none-zero probabilities, then use a binary mask to filter other less relevant clusters and obtain the final item-cluster affiliation matrix $\Mat{X}$. However, due to the non-differentiability and discrete nature of $\argtop{}$ operation, the computation is intractable thus cannot be optimized with gradients. Therefore, we relax it by tempered softmax:
{\setlength{\abovedisplayskip}{3pt}
\setlength{\belowdisplayskip}{3pt}
\begin{gather}\label{equ:topk-trick}
    m_{ic} \approx \tilde{m}_{ic} = \frac{\exp({\cos (\Mat{v}_i, \Mat{h}_c)/T})}{\sum\nolimits_c\exp(\cos (\Mat{v}_i, \Mat{h}_c)/T)},
\end{gather}}where $T$ is a hyperparameter called \textit{temperature}, which controls the entropy of the distribution. Thus, for back-propagation, the continuous relaxation $\tilde{m}_{ic}$ is used to approximate $m_{ic}$. However, when calculating the probability of taste clusters, we should directly use $m_{ic}$ instead of $\tilde{m}_{ic}$, to be consistent with the real affiliation relationship. To close the gap between forward pass and backward pass, we follow a similar idea to reparameterized trick~\cite{iclr17gumbel,zhang2022gateformer}, and rewrite $m_{ic}$ as follows:
{\setlength{\abovedisplayskip}{3pt}
\setlength{\belowdisplayskip}{3pt}
\begin{gather}\label{equ:detach-gradient}
    \hat{m}_{ic} = \tilde{m}_{ic}  + \mathrm{detach\_gradient} (m_{ic} - \tilde{m}_{ic}),
\end{gather}}where the detach\_gradient will prevent the gradient from
back-propagating through it. In the forward pass, detach\_gradient has no effect, thus the affiliated clusters can be directly computed by $\argtop{}$ operation. In the backward pass,  detach\_gradient takes effect, so $\nabla_{\hat{m}_{ic}} \Lapl = \nabla_{\tilde{m}_{ic}} \Lapl$. So the whole computation is fully differentiable and can be smoothly optimized with the model.

We use the same mechanism to obtain the affiliations between users and clusters. Specifically, the user-cluster similarity matrix can be computed as $\tilde{\Mat{A}} = \Mat{Y} \times \tilde{\Mat{X}}$, where $\Mat{Y}$ is the user-item interaction matrix. Then, we also force each user to connect with Top-$n$ taste clusters to derive the user-cluster affiliation matrix $\Mat{A}$. 

\vspace{3pt}
\noindent
\textbf{Forest Mechanism.}
Preference sparsification with taste clusters is an approximation of latent factor-based collaborative filtering. As a result, ECF inevitably suffers from sub-optimal performance and limited diversity.To overcome these two challenges, we introduce a straightforward yet impactful ensemble approach that improves the expressiveness of taste clusters and preserves its interpretability. Specifically, for each individual ECF model, we randomly select $|\Set{C}|$ items as the initial taste clusters and use different random seeds for model training. Then we train $F$ different ECF instances to form the final ECF model, and the final prediction is based on the summation of all $F$ models. We find this simple random mechanism can boost the performance of ECF and  provide a comprehensive explanation for predictions (the impact of different number of models is discussed in Section~\ref{sec:exp-forest}).

\subsection{Optimization of ECF}
Next, we introduce an end-to-end optimization method to train ECF, with the aim of constraining the taste cluster from different perspectives to meet its definition. First, since taste clusters can reveal both the user's interests and the similarity between items, it is reasonable to directly predict user preference for an item through their common taste clusters (as described in Equation (\ref{equ:ecf-prediction})). Therefore, the cluster-based predictions can be optimized with BPR loss~\cite{uai09BPRloss}:
{\setlength{\abovedisplayskip}{3pt}
\setlength{\belowdisplayskip}{3pt}
\begin{gather}\label{equ:sparse-bpr}
    \Lapl_\text{CS} = \sum\nolimits_{(u, i, j) \in \Set{O}}-\ln \sigma(\hat{y}_{u i}-\hat{y}_{u j}),
\end{gather}}where $\Set{O} = \{(u,i,j) | y_{ui} = 1, y_{uj} = 0\}$ denotes the training set of user-item interactions. Then we consider the informativeness of tags. We first calculate the tag distribution of taste clusters:
{\setlength{\abovedisplayskip}{3pt}
\setlength{\belowdisplayskip}{3pt}
\begin{gather}\label{equ:tag-distribution}
    \tilde{\Mat{D}} =  \Mat{X}^{\top} \Mat{E},
\end{gather}}where $\Mat{E} \in \Space{R}^{|\Set{I}| \times |\Set{T}|}$ is the multi-hot matrix where each entry $e_{it}$ denotes whether item $i$ has the tag $t$, and $\tilde{\Mat{D}}$ represents the tag frequency of each taste cluster. Intuitively, we can directly select the most frequent tags as the description of taste clusters. However, this naive approach would result in the indistinguishability between different clusters, since some tags are commonly shared by most items, and they would dominate the tag distribution across different taste clusters. Take the game genres at Xbox dataset as an example. Most of the games are online and allow for mutiplayer, thus selecting the ``Multiplayer-Online'' tag fails to provide informative description for taste clusters. To tackle this problem, we first reweight each tag according to its frequency of occurrence in items, and then compute the weighted distribution:
{\setlength{\abovedisplayskip}{3pt}
\setlength{\belowdisplayskip}{3pt}
\begin{gather}\label{equ:tag-tfidf}
    d_{ct} =  \tilde{d}_{ct} \times \log (\frac{N}{f_t + \epsilon}),
\end{gather}}where $N$ is the number of items, $f_t$ is the frequency of tag $t$ across all items, and $\epsilon$ is a small fraction (we set it as $10^{-6}$) to avoid numerical overflow. Thus, the above equation considers both tag frequency and items' affiliations for tag informativeness: the more frequently the distinctive tag appears, the more informative it is for interpreting the taste cluster.

To improve the understanding of taste clusters, we advocate for an appropriate number of tags, not too few or too many. An excessive or inadequate number of tags makes it difficult to identify users' true interests and results in a complex or abstract interpretation of the taste cluster." To achieve that, we first  normalize their frequencies as:
{\setlength{\abovedisplayskip}{3pt}
\setlength{\belowdisplayskip}{3pt}
\begin{gather}\label{equ:tag-softmax}
    \beta_{ct} =  \frac{\exp ( d_{ct}/\tau)}{\sum\nolimits_{j\in\Set{T}}\exp ( d_{cj}/\tau)},
\end{gather}}where $\tau$ is the is the temperature hyperparameter. At low temperatures, the distribution sharpens, causing only the tag with the highest score to stand out. As the temperature increases, the distribution becomes more even and the distinctiveness between tags will decrease. Then, we consider maximizing the likelihood of the probabilities of Top-$P$ tags so that the taste clusters can be easily interpreted by those tags:
{\setlength{\abovedisplayskip}{3pt}
\setlength{\belowdisplayskip}{3pt}
\begin{gather}\label{equ:tag-likelihood}
    \Lapl_\text{TS} = \sum\nolimits_{c \in \Set{C}} \sum\nolimits_{t\in \argtop{P}(\beta_{c})}-\log \beta_{ct},
\end{gather}}where $\Set{C}$ denotes the set of taste clusters. We fix the number of tags $P=4$ for all experiments, since it achieves a good balance between informativeness and readability.

Moreover, different taste clusters should contain different items and reveal different user preferences, so that the latent interest space could be better preserved. Hence, for better model capacity and diversity, we encourage the embeddings of taste clusters to differ from each other's. There are many methods for independence modeling, such as distance correlation~\cite{07pearson}, orthogonality~\cite{ijcai19hifi}, and mutual information~\cite{icml18mutual,www21kgin}. Here we opt for mutual information for all experiments due to its simplicity and effectiveness:
{\setlength{\abovedisplayskip}{3pt}
\setlength{\belowdisplayskip}{3pt}
\begin{gather}\label{equ:mutual-information}
\Lapl_\text{IND} = \sum\nolimits_{c \in C} -\log \frac{\exp(s(\Mat{h}_c,\Mat{h}_c))}{{\sum\nolimits_{c^\prime \in C}\exp(s(\Mat{h}_c,\Mat{h}_{c^\prime}))}},
\end{gather}}where $s(\cdot)$ is the similarity function to measure the associations of any two taste clusters, which is set as cosine function here. Finally, we consider learning taste clusters by putting the above objective functions together: $\Lapl_\text{TC} = \Lapl_\text{CS} + \Lapl_\text{TS} + \Lapl_\text{IND}$.
For simplicity we do not need to tune any weighting coefficients for each loss. However, due to the $\argtop{}$ operations in affiliation selections, the supervised signals are sparse and hard to converge. Thus, we add auxiliary supervised signals from user-item predictions:
{\setlength{\abovedisplayskip}{3pt}
\setlength{\belowdisplayskip}{3pt}
\begin{gather}\label{equ:dense-loss}
\Lapl_\text{CF} = \sum\nolimits_{(u, i, j) \in \Set{O}}-\ln \sigma(\Mat{u}_u^{\top} \Mat{v}_i - \Mat{u}_u^{\top} \Mat{v}_j),
\end{gather}}where $\Mat{u}_u$ denotes the embeddings of user $u$. We choose inner product of embeddings to measure the similarity between users and items for simplicity, but more sophisticated embedding-based models (\eg LightGCN~\cite{sigir20lighGCN}) can also be applied, which will be discussed in Section~\ref{sec:further-study}. By combing the taste clusters loss and collaborative loss, we minimize the following objective function to learn the model parameters:
{\setlength{\abovedisplayskip}{3pt}
\setlength{\belowdisplayskip}{3pt}
\begin{gather}\label{equ:ecf-loss}
\Lapl_\text{ECF} = \Lapl_\text{TC} + \lambda \Lapl_\text{CF},
\end{gather}}where $\lambda$ a hyperparameter to control the impact of auxiliary collaborative signals.

\subsection{New Metrics for Explainability}\label{sec:explanation-metrics}

Here we aim to design a holistic set of quantitative metrics to qualify the effectiveness of explanation \wrt taste clusters:

\begin{itemize}[leftmargin=*]
    \item \textbf{In-cluster item coverage} denotes the average proportion of items in a taste cluster that the selected tags can cover. With a slight abuse of notation, we use $\Set{T}_i$ to denote the tags of item $i$ and $\Set{T}_c$ to denote the tags of cluster $c$:
    {\setlength{\abovedisplayskip}{3pt}
\setlength{\belowdisplayskip}{3pt}
    \begin{gather}\label{equ:coverage}
    \text{Cov.} = \frac{1}{|\Set{C}|} \sum\nolimits_{c\in\Set{C}} \sum\nolimits_{i\in c} \frac{\mathbbm{1}( \left| \Set{T}_i\cap \Set{T}_{c}  \right|)}{|c|},
    \end{gather}}where $\mathbbm{1}(x) = 1$ if and only if $x>0$, otherwise $0$. When the item coverage ratio is high, we deem that these tags can be properly used as the descriptive interpretation for the taste cluster.
    \item \textbf{Tag utilization} represents how many unique tags are used for interpreting taste clusters:
    {\setlength{\abovedisplayskip}{3pt}
    \setlength{\belowdisplayskip}{3pt}
    \begin{gather}\label{equ:utilization}
    \text{Util.} = \frac{1}{|\Set{T}|} \bigcup_{c\in\Set{C}} \Set{T}_{c}, 
    \end{gather}}where we union all the selected tags from each taste cluster, and a higher tag utilization indicates a more diverse interpretation of interest patterns.
    \item \textbf{Silhouette}~\cite{ROUSSEEUW198753} is a clustering metric which measures the similarity difference between intra-cluster and inter-cluster items:
    {\setlength{\abovedisplayskip}{3pt}
    \setlength{\belowdisplayskip}{3pt}
    \begin{gather}\label{equ:silhouette}
    \text{Sil.} = \frac{1}{|\Set{C}|^2}\sum\nolimits_{c_1\in \Set{C}} \sum\nolimits_{c_2\in \Set{C}}
    \frac{b(c_1,c_2) - a(c_1)}{\max \{a(c_1), b(c_1,c_2)\}},
    \end{gather}}where $b(c_1,c_2)$ is the mean cosine similarity between all disjoint items in $c_1$ and $c_2$, and $a(c_1)$ is the mean cosine similarity of items in taste cluster $c_1$. The high silhouette indicates the items are similar in the same taste clusters.
    \item \textbf{Informativeness} measures the distinctiveness of selected tags to represent the items in the taste cluster:
    {\setlength{\abovedisplayskip}{3pt}
    \setlength{\belowdisplayskip}{3pt}
    \begin{gather}\label{equ:informativeness}
    \text{Info.} = \frac{1}{|\Set{C}|} \sum\nolimits_{c_i \in \Set{C}} \frac{|R(\Set{T}_{c})\cap c|}{|c|},
    \end{gather}}where $R(\Set{T}_{c})$ is an post-hoc discriminator that predicts the most likely top $|c|$ items given the tags of taste cluster $c$ . We detail the implementation of the function $R(\cdot)$ in Appendix~\ref{sec:appendix-predictor}. A greater informativeness score implies that those tags more accurately characterize the items within the taste clusters.
\end{itemize}

The above four metrics measure the quality of taste clusters from different aspects. Since ECF depends on taste clusters to make recommendation and explanation, these metrics can be also viewed as the evaluation of explanation. We further provide an overall metric by normalizing each metric with a random approach. In addition, since explainability is a human notion and cannot be fully evaluated computationally without input from human users, we also included two user studies as complementary to the metrics, which will be detailed in Section~\ref{sec:eval-explanation}.

\subsection{Complexity Analysis}

\subsubsection{\textbf{Model Size}} 
We analyze the size of ECF from both training and inference perspectives. During the training, the model parameters of single ECF consist of (1) ID embedding of users and items $\{\Mat{U},\Mat{V} | \Mat{U}\in \Space{R}^{|\Set{U}|\times d},\Mat{V}\in\Space{R}^{|\Set{I}|\times d} \}$, which are also used by all embedding-based methods; and (2) ID embedding of taste clusters $\{\Mat{H}| \Mat{H}\in\Space{R}^{|\Set{C}|\times d} \}$. We utilize $F$ different ECF models for training, so that the overall parameters are $F$ times of the single model parameters. During the inference, different from other embedding-based methods, ECF only needs the sparse affiliations between users/items and clusters for prediction, and the parameters are $F (m |\Set{I}| +n |\Set{U}|)$, where $m$ and $n$ are the number of affiliations. The size of ECF is similar or even less than typical embedding-based methods (\ie MF need $d (|\Set{I}|+|\Set{U}|)$ parameters) when selected affiliations $m, n \ll d$.

\subsubsection{\textbf{Time Complexity}}
Time cost of ECF mainly comes from the taste clusters learning. For collaborative similarity learning of taste clusters, the computational complexity of calculating the affiliations is $O(|\Set{C}|(d|\Set{I}| + |\Mat{Y}|)$, where $|\Set{C}|$, $|\Set{I}|$, $|\Mat{Y}|$ and $d$ denote the number of taste clusters, items, interactions, and the embedding size. For tag similarity learning, the computational complexity of tag distribution is $O(|\Set{C}||\Set{I}||\bar{t}|)$, where $|\bar{t}|$ is the average number of tags that items have. As for independence modeling, the cost of mutual information is $O(d|\Set{C}|^2)$. Besides, computing the auxiliary collaborative signals would cost $O(d|\Set{I}||\Set{U}|)$. Thus, the time complexity of the whole training epoch is $O(|\Set{C}||\Mat{Y}| + d|\Set{I}||\Set{U}|)$ when the number of taste clusters is far more less than the number of users/items. Under the same experimental settings
(\ie same embeddings size and same number of tags), ECF has comparable complexity to EFM and AMCF, two representative explainable methods.



%% file: 3_exp.tex
\section{Experiments}
We design experiments to answer the following research questions: 
(1) How does ECF perform in accuracy, compared with related competitive methods? 
(2) What is the quality of ECF's interpretability?
(3) What is the impact of various components and hyperparameters on the effectiveness of ECF? 
(4) What do the generated taste clusters resemble and what are the potential uses for ECF?

\subsection{Experimental Settings}

\begin{table}[t]
    \centering
    \vspace{-10pt}
    \caption{Statistics of the datasets used in our experiments.}
    \vspace{-10pt}
    \label{tab:dataset}
    \resizebox{0.44\textwidth}{!}{
    \begin{tabular}{c|r|r|r|r}
    \hline
    Dataset          & \#Users & \#Items & \#Interactions & \#Tags \\ \hline\hline
    Xbox      & 465,258  & 330  & 6,240,251      & 115 \\
    MovieLens & 6,033 & 3,378  & 836,434  & 18      \\
    Last-FM & 53,486 & 2,062  & 2,228,949      & 54 \\\hline
    \end{tabular}}
    \vspace{-10pt}
\end{table}

We use three real-world datasets for experiments: Xbox,  MovieLens and Last-FM, which vary in domain, size, tag numbers, and sparsity. The Xbox dataset is provided by Microsoft Gaming and collected from the GamePass scenario~\footnote{\url{https://www.xbox.com/en-US/xbox-game-pass}}. GamePass is a popular membership-based service offered by Xbox, with which users can freely play a few hundreds of high-quality games (that is why in Table~\ref{tab:dataset} Xbox has only 330 items). MovieLens is a widely used dataset which contains user rating history and the types of movies, we follow~\cite{ijcai20amcf} to get 18 neat tags for movie description. Last-FM is a public music listening dataset. We use the official APIs~\footnote{\url{https://www.last.fm/api}} of Last.fm to obtain the tags for each track and discard rare tags which are annotated by less than 50 people. Besides, to reduce noise, we adopt the 10-core setting, \ie retaining users and items with at least ten interactions. We use the same data partition with previous study~\cite{ijcai20amcf} for comparison (i.e., the proportions of training, validation, and testing set are 80\%, 10\%, and 10\% for all datasets). Table~\ref{tab:dataset} summarizes the basic statistics of the three datasets. 

\subsection{Evaluation of Accuracy (RQ1)}\label{sec:exp-performance}

\subsubsection{\textbf{Baselines}}
We compare ECF with two groups of models:
\begin{itemize}[leftmargin=*]
    \item \textbf{MF}~\cite{uai09BPRloss}, \textbf{NCF}~\cite{www17ncf},  \textbf{CDAE}~\cite{wsdm16cdae} and \textbf{LightGCN}~\cite{sigir20lighGCN} are four strong and representative embedding-based CF methods which learn users' hidden preference from interactions. But they suffer from poor explainability.
    \item \textbf{EFM}~\cite{sigir14efm} and \textbf{AMCF}~\cite{ijcai20amcf} are two strong explainable recommendation models, but they fail to satisfy all three explanation merits. We adapt EFM model slightly to make it fit for our scenario (see Appendix~\ref{sec:appendix-efm} for details).
\end{itemize}
We also add the single model version of ECF for variant comparison, denoted as ``ECF$_{single}$''.

\subsubsection{\textbf{Evaluation Metrics}}
We adopt two widely used evaluation protocols~\cite{uai09BPRloss,sigir14efm,www21kgin} for top-$K$ recommendation: Recall@\textit{K} and NDCG@\textit{K}. Hyper-parameters are reported in Appendix~\ref{sec:appendix-hyerparameter}.

\subsubsection{\textbf{Performance Comparison}}
The performance \wrt Recall@20 and NDCG@20 is reported in Table~\ref{tab:top-k-recommendation}. More experiments with different $K$ can be found in Appendix~\ref{sec:appendix-topk}. In comparison, embedding-based methods achieve competitive performance on all datasets.  LightGCN outperforms all other baselines on MovieLens and Last-FM datasets, we contribute this to the ability of capturing the high-order user-item relationship in bipartite graph. On the other hand, explainable methods - EFM and AMCF - only show a slight improvement over MF. As for the ECF$_{single}$, we can see the performance drops compared with MF, due to the sparsity design of taste clusters affiliations for interpretability. However, the forest mechanism in ECF makes up for its shortcomings in accuracy and outperforms all the baselines by a large margin (including the forest version of MF, see Appendix~\ref{sec:appendix-topk} for details). 
This demonstrates that ECF is not taking a trade-off between accuracy and interpretability, instead, it can achieve excellent accuracy performance while providing interpretability.

\begin{table}[t]
    \vspace{-10pt}
    \caption{Top-$20$ recommendation results. ``${\dagger}$'' indicates the
improvement of the ECF over the baseline is significant at the level of 0.05.}
    \centering
    \vspace{-10pt}
    \label{tab:top-k-recommendation}
    \resizebox{0.49\textwidth}{!}{
    \begin{tabular}{c|c c |c c| c c}
    \hline
    \multicolumn{1}{c|}{\multirow{2}*{}}&
    \multicolumn{2}{c|}{Xbox} &
    \multicolumn{2}{c|}{MovieLens} &
    \multicolumn{2}{c}{Last-FM} \\
      &Recall & NDCG & Recall & NDCG & Recall & NDCG\\
    \hline
    MF  & 0.5048 & 0.3268 & 0.1603 & 0.2416 & 0.0658 & 0.0506 \\
    NCF & 0.4746 & 0.2931 & 0.1606 & 0.2406 & 0.0618 & 0.0401 \\
    CDAE & \underline{0.5192} & \underline{0.3286} & 0.1627 & 0.2499 & 0.0589 & 0.0534 \\
    LightGCN & 0.4933 & 0.3261 & \underline{0.1854} & \underline{0.2698} & \underline{0.0788} & \underline{0.0675} \\ \hline
    EFM & 0.5070 & 0.3312 & 0.1702 & 0.2525 & 0.0703 & 0.0549 \\
    AMCF & 0.5036 & 0.3217 & 0.1604 & 0.2405 & 0.0675 & 0.0516 \\  \hline
    ECF$_{single}$ & 0.4231 & 0.2331 & 0.1068 & 0.1501 & 0.0467 & 0.0380 \\
    ECF & \textbf{0.5922}$^{\dagger}$  & \textbf{0.3721}$^{\dagger}$  & \textbf{0.2124}$^{\dagger}$  & \textbf{0.2903}$^{\dagger}$  & \textbf{0.0851}$^{\dagger}$  & \textbf{0.0773}$^{\dagger}$ \\
    \hline
    \end{tabular}}
    \vspace{-10pt}
\end{table}

\subsection{Evaluation of Explainability (RQ2)} \label{sec:eval-explanation}
Then we evaluate the explainability of ECF. As elaborated in Section~\ref{sec:explanation-metrics}, we utilize four metrics to evaluate the explainability of our model: in-cluster item coverage, tag utilization, silhouette and informativeness. Since ECF is a cluster-based CF method, for comparison, we use three strong competitors to construct the clusters from different perspectives as baselines. Specifically, \textbf{TagCluster} is a tag-oriented method which groups items with the same tags into clusters; \textbf{K-means} is a similarity-oriented method which utilizes item embedding from MF to perform K-means algorithm; \textbf{Random} is a baseline method by randomly selecting items into clusters. The detailed implementation of baselines is reported in Appendix~\ref{sec:appendix-baselines}.

To have an overall understanding of the explainability, we introduce an aggregated metric denoted as \textsl{Overall} in Table~\ref{tab:explanation}. 
To overcome the scale inconsistency issue in different metrics, we first normalize each metric by calculating the relative gain over the \textsl{Random} baseline, then sum up the normalized four scores as the \textsl{Overall} metric. 
In terms of the overall metric, ECF outperforms all competitors by a large margin. Specifically, all the baselines can only preserve a certain aspect while ignoring the rest three aspects. For instance, since TagCluster only considers the explicit tags for clustering, it fails to preserve the collaborative filtering-related similarity in the cluster. Furthermore, traditional clustering methods like K-means suffer from low coverage ratio and poor informativeness, revealing their inability to form significant taste clusters with descriptive tags. In contrast, ECF considers all relevant factors, thus preventing significant shortcomings in any one metric.

Besides, we have included two user studies on the LastFM dataset as complementary to Table~\ref{tab:explanation}. These studies involved 30 volunteers evaluating the explainability of both taste clusters and user-item recommendations. Volunteers were asked to participate two tasks as human evaluation. The task description and results are detailed in Appendix~\ref{sec:appendix-human}, and ECF also achieved the best among baselines from human judgements. 

\begin{table}[t]
\centering
\vspace{-10pt}
\caption{Explainability Evaluation of ECF. 
}
\vspace{-10pt}
\label{tab:explanation}
\resizebox{0.45\textwidth}{!}{
\begin{tabular}{c|c|c|c|c|c}
\hline

 \multicolumn{1}{c|}{\textbf{Method}} & \multicolumn{1}{c|}{\textbf{Cov.}} & \multicolumn{1}{c|}{\textbf{Util.}} &
\multicolumn{1}{c|}{\textbf{Sil.}} & \multicolumn{1}{c|}{\textbf{Info.}}& \multicolumn{1}{c}{\textbf{Overall}}  \\ \hline
\multicolumn{6}{c}{Xbox}  \\  \hline
 ECF   & \underline{0.8002} & \textbf{0.7052}	& \underline{0.2604}	& \textbf{0.3162}	& \textbf{1.7463}\\
                                      TagCluster   & \textbf{0.9950} & 0.2878		&-0.1788&\underline{0.1579}	 & 0.9262 \\
                                     K-means   &0.5710 & \underline{0.3739}		&\textbf{0.4286}	&0.0185	  & \underline{1.0563}\\
                                      Random   &0.5396 & 0.1450		&-0.3614	&0.0125	 & 0.0000 \\
                                      \hline
\multicolumn{6}{c}{MovieLens}  \\  \hline
 ECF   & \underline{0.7992} & \textbf{0.7778}		& \underline{0.1964}	&\textbf{0.3131}& \textbf{1.5651}	\\
                                    TagCluster   &\textbf{0.991} & \underline{0.5259} &-0.2573	&\underline{0.1517}	 & 0.8898 \\
                                      K-means   &0.6877 & 0.4478	&\textbf{0.3265}	&0.0168	 & \underline{0.9573} \\
                                      Random   &0.5933 & 0.3672	&-0.4452	&0.0061	 & 0.0000 \\ 
                                      \hline
\multicolumn{6}{c}{Last-FM}  \\  \hline
 ECF   &\underline{0.7648} & \textbf{0.6259}		&\underline{0.1584}	&\textbf{0.2996}& \textbf{1.5352}	\\
                                    TagCluster   &\textbf{0.9880} & 0.3703		&-0.2511	&\underline{0.1206}	  & 0.9143\\
                                     K-means &0.5667  & \underline{0.4841}		&\textbf{0.3197}	&0.0182	 & \underline{1.0752} \\
                                       Random   &0.5385 & 0.2275	&-0.4673	&0.0148	 & 0.0000\\
                                      \hline
\end{tabular}}
\vspace{-5pt}
\end{table}





\begin{figure}
    \centering
    \includegraphics[width=0.23\textwidth]{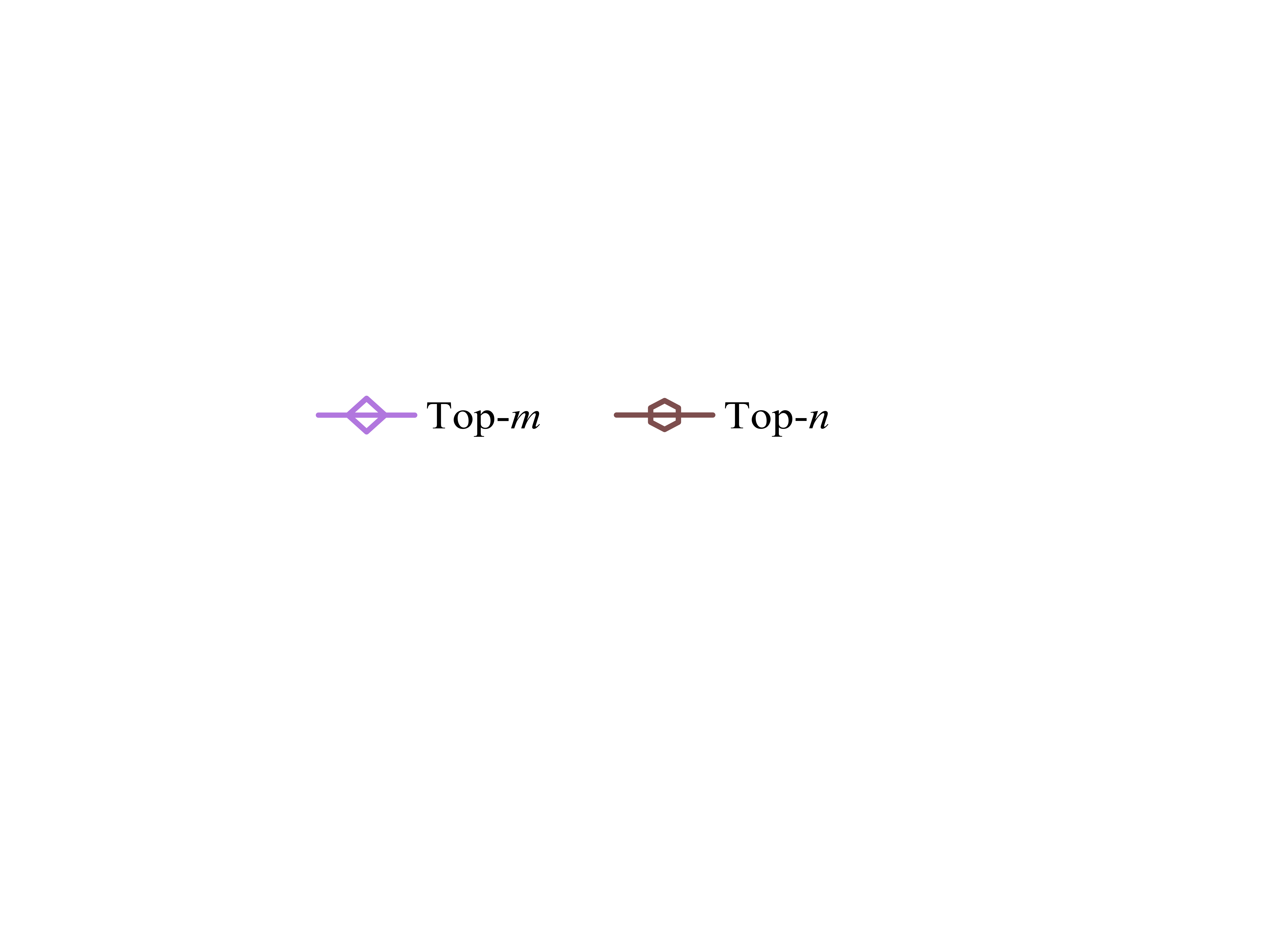}
    \includegraphics[width=0.44\textwidth]{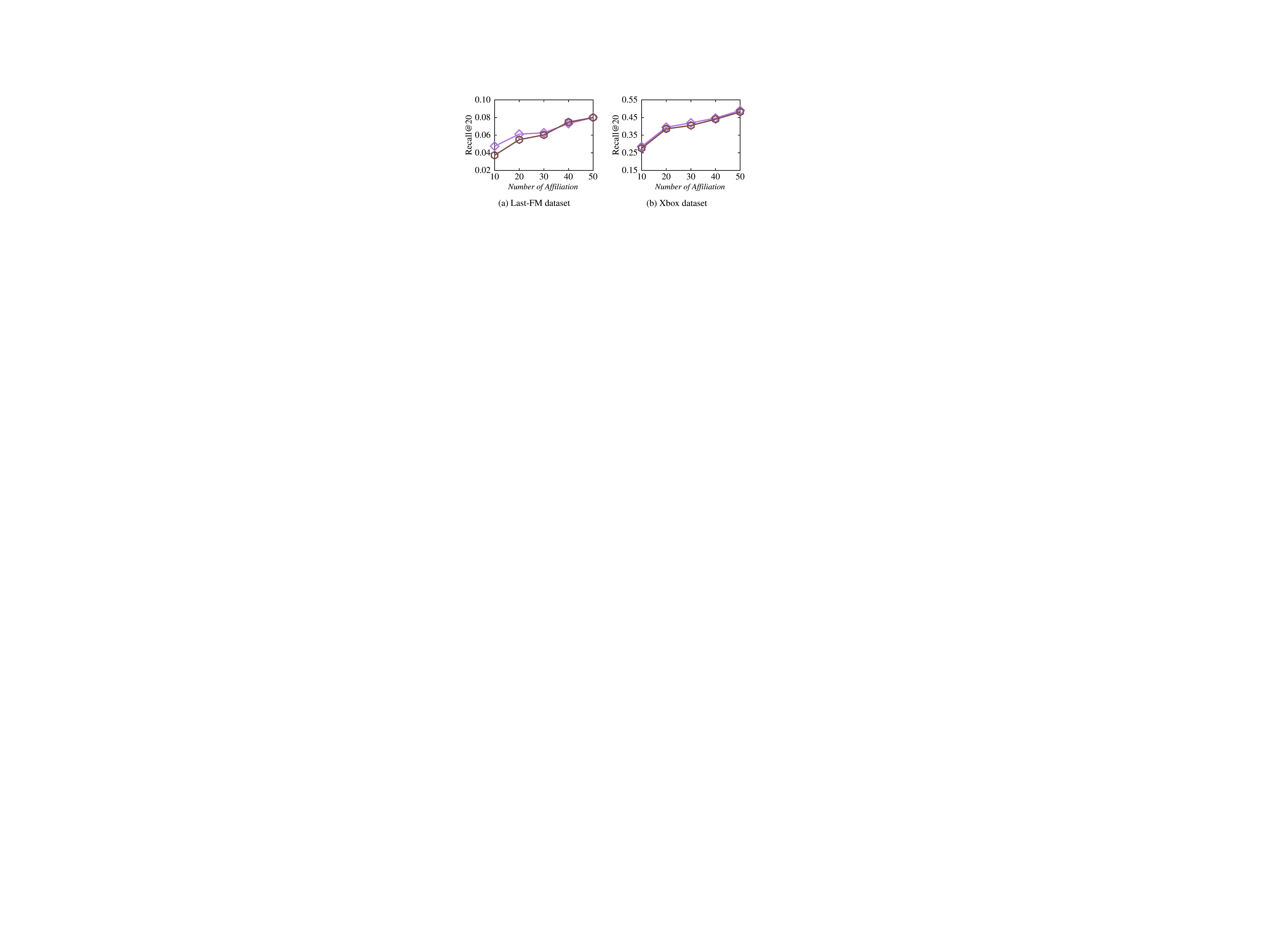}
    \vspace{-3mm}
    \caption{Impact of affiliation number on items (Top-$m$) and on users (Top-$n$).}
    \label{fig:impact-of-topk}
    \vspace{-3mm}
\end{figure}

\begin{figure*}[t]
	\centering
	\vspace{-15pt}
	\includegraphics[width=0.98\textwidth]{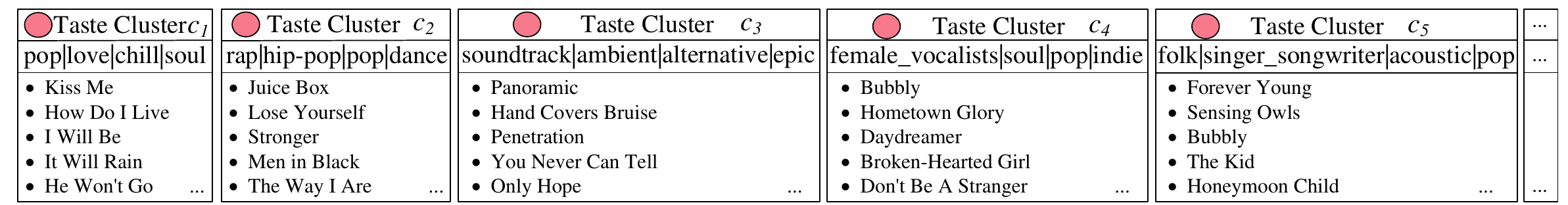}
	\vspace{-10pt}
	\caption{5 real examples of learned taste clusters on Last-FM dataset.}
	\label{fig:casestudy-tastecluster}
\end{figure*}

\subsection{Study of ECF (RQ3)} 

\subsubsection{\textbf{Impact of Top-$m$ and Top-$n$}} 
As described in Section~\ref{sec:framework}, we use Top-$m$ and Top-$n$ to select the most relevant taste clusters for items/users. The experimental results of different $m$ and $n$ on Xbox and Last-FM dataset are reported in Figure~\ref{fig:impact-of-topk}, while the results on MovieLens dataset is omitted due to similar trend and limited space. We observe that the recommendation performance boosts dramatically when $m$ and $n$ increase, especially from 10 to 20. This is because when the affiliations between taste clusters and users/items are sparse, it is unable to model the complex and latent users' preference, as well as the multi-faceted semantics of items. However, large $m$ and $n$ would make the explanation results hard to understand because of the excessive number of reachable paths between users, taste clusters and items. Therefore, to balance the readability and accuracy, we set $m$ and $n$ to 20 for all datasets.

\begin{figure}[t]
	\centering
    \includegraphics[width=0.32\textwidth]{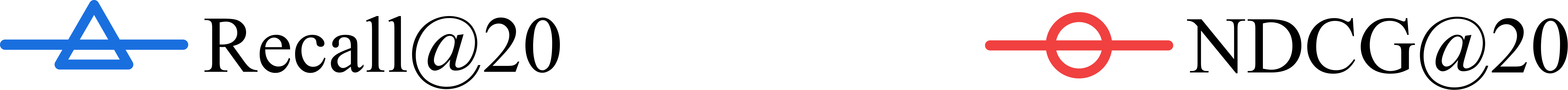}
	\subfigure{
        \includegraphics[width=0.22\textwidth]{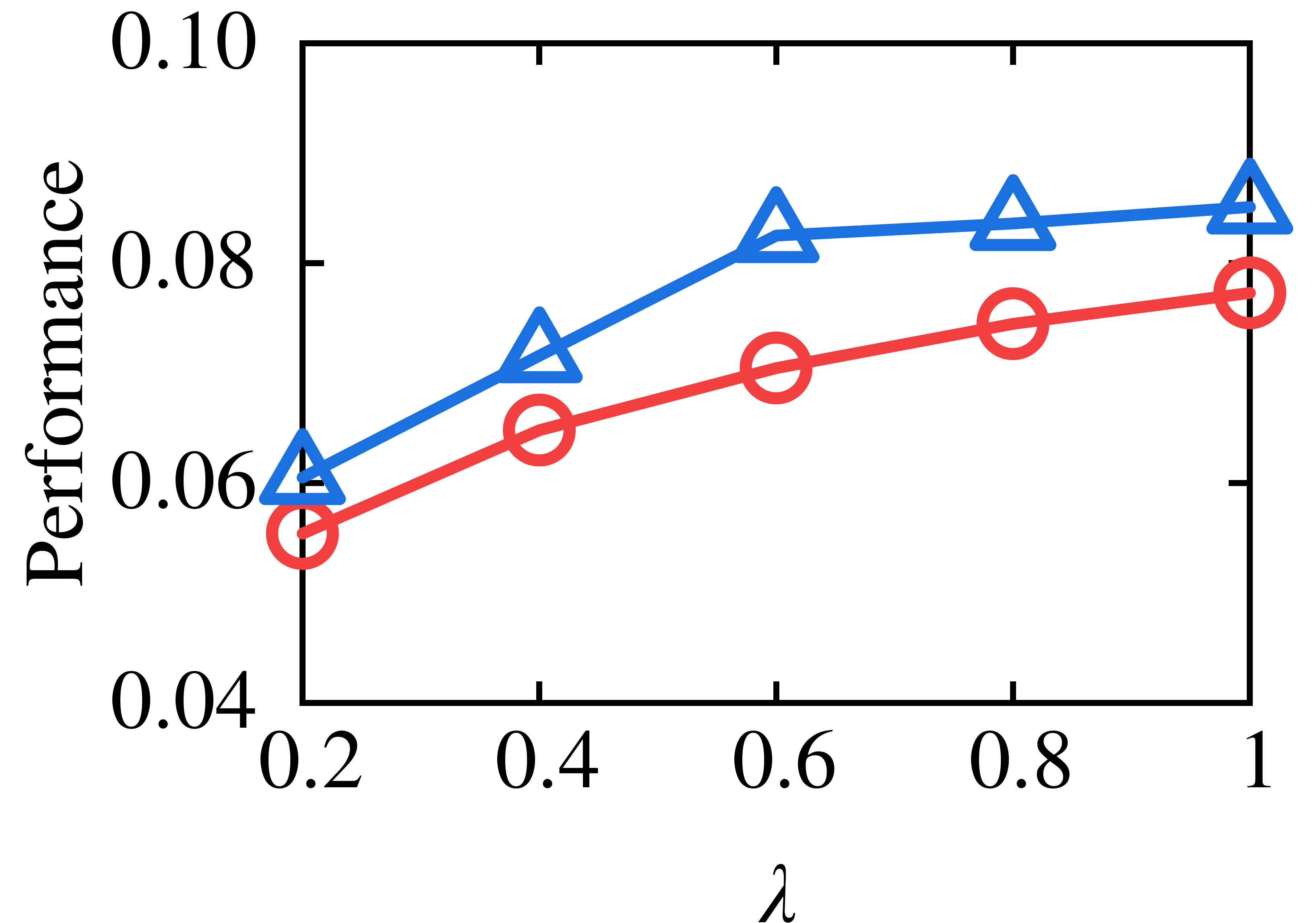}}
	\hspace{1.4mm}
	\hspace{-0.25cm}
	\subfigure{
		\includegraphics[width=0.22\textwidth]{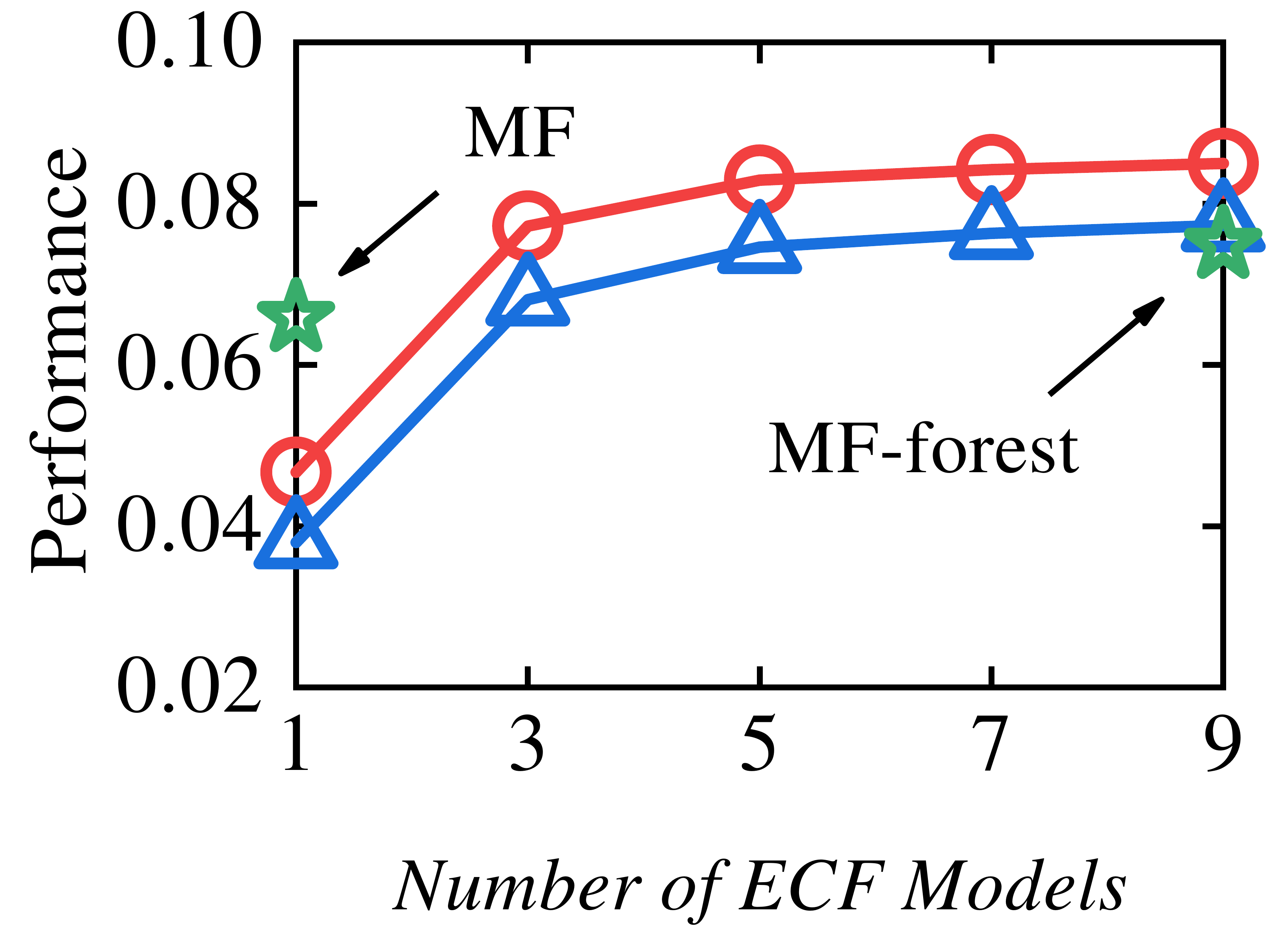}}
	\vspace{-5mm}
	\caption{Hyperparameter study on Last-FM dataset. Left: impact of $\lambda$ in Eq.(\ref{equ:ecf-loss}). Right: impact of the forest mechanism.}
	\label{fig:impact-of-lambda-forest}
\end{figure}

\subsubsection{\textbf{Impact of $\lambda$}}
We then verify the effectiveness of the auxiliary collaborative signals. The results on Last-FM dataset is illustrated in the left of Figure~\ref{fig:impact-of-lambda-forest}, and the results on other two datasets are reported in Appendix~\ref{sec:appendix-impact-forest}. We observe that a small $\lambda$ value results in a significant decline in the performance of ECF, due to the insufficient supervised signals for the learning of taste clusters. On the other hand, when $\lambda$ exceeds 0.6, the improvement is minimal, indicating that ECF can be optimized with adequate collaborative signals. Thus, we fix $\lambda$ to 0.6 for all of our experiments.

\subsubsection{\textbf{Impact of Forest Mechanism}}\label{sec:exp-forest}
To analyze the effect of the ensemble version of ECF, we gradually increase the size of forest, and the performance results are reported in the right of Figure~\ref{fig:impact-of-lambda-forest}. We find that there is a notable performance gain of ECF, compared with the single ECF model. We contribute the improvement to the holistic modeling of users' latent interest space with diverse taste clusters. Besides, we also find that the performance of ECF is even better than the forest version of MF (see Figure~~\ref{fig:impact-of-lambda-forest} and Appendix~\ref{tab:appendix-top-k-recommendation} for details), which indicates the effectiveness of taste cluster learning. 

\subsubsection{\textbf{Flexibility of ECF}}\label{sec:further-study}  
Since the ECF framework only relies on embeddings to learn explainable taste clusters, it can be viewed as a model-independent explainable paradigm, which is easily applied with other popular embedding-based methods. Thus, we also incorporate ECF with LightGCN~\cite{sigir20lighGCN}, a competitive graph-based embedding method, to investigate its flexibility. The results in Figure~\ref{sec:appendix-ECF-LightGCN} indicate that the performance and explainability of ECF would also benefit from more complex embedding-based methods, which demonstrates the superiority and universality of ECF (see Appendix~\ref{sec:appendix-ECF-LightGCN} for details).

\subsection{Case Study and Applications (RQ4)}
In this section, we present examples from the Last-FM dataset to provide an intuitive understanding of taste clusters and demonstrate the explainability of ECF. Additionally, we showcase the versatility of ECF by highlighting useful scenarios beyond item recommendations.

\subsubsection{\textbf{Learned Taste Clusters}}

We first illustrate the generated five taste clusters in Figure~\ref{fig:casestudy-tastecluster}. Each taste cluster is interpreted with four tags, which are the best description of the tracks within. For instance, all the tracks in taste cluster $c_3$ are soundtrack, while some are ambient music (\eg ``Hand Covers Bruise''), and some are tagged with ``alternative'' (\eg ``Only Hope''). Besides, each taste cluster is able to explicitly represent certain users' preference. Taking the taste cluster $c_2$ as an example, users who have a high affiliation score with it tend to favor rap and hip-pop music, especially these tracks that are accompanied with dancing. Moreover, different taste clusters have different tags (\eg pop music in $c_1$ and folk music in $c_5$), indicating the diversity of users' preference.

As for the relationship between taste clusters and  items, we find taste clusters have the ability to serve as complementary information to the tagging of items, which we call \textit{tag discovery}. For instance, folk song ``Bubbly'' in taste cluster $c_5$ is tagged with ``female\_vocalists|pop|folk|acoustic|love'' by Last.fm. Thus, we can suspect whether ``Bubbly'' has the missing tag ``singer\_songwriter'' since other tags in the taste cluster $c_5$ (\ie ``folk|acoustic|pop'') are perfectly to describe the track. In fact, the singer of the song, Colbie Caillat, is also a songwriter who wrote the song. Therefore, taste clusters can be useful for automatic tag completion.

\begin{figure}[t]
    \centering
    \includegraphics[width=0.42\textwidth]{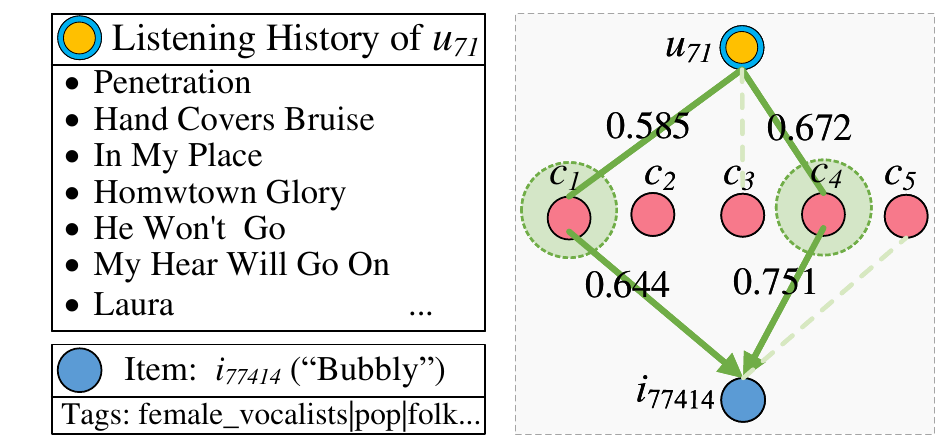}
    \vspace{-5pt}
    \caption{Explanations of the recommendation of user $u_{71}$ to track $i_{77414}$ (``Bubbly'') on Last-FM dataset. $c$ denotes the taste cluster which shows in Figure~\ref{fig:casestudy-tastecluster}. The tags of the track $i_{77414}$ are ``female\_vocalists|pop|folk|acoustic|love''.}
    \label{fig:casestudy-rec}
\end{figure}

\subsubsection{\textbf{Recommendation Explanation}}\label{exp-case-study-rec}
Next, we demonstrate the explanation ability of ECF. Taking the recommendation of user $u_{71}$ to track $i_{77414}$ (``Bubbly'') as an example, Figure~\ref{fig:casestudy-rec} shows the listening history of $u_{71}$ as well as the recommendation logic of ECF, where we have the following observations.

First, ECF maps users as well as items with their corresponding taste clusters. For user $u_{71}$, since he/she has listened different kinds of songs from soundtrack to pop and soul music, three different taste clusters (\ie $c_1$, $c_3$ and $c_4$.) have been affiliated to reveal his interest. Similarly, track $i_{77414}$ is also connected with three taste clusters (\ie $c_1$, $c_4$ and $c_5$), which is able to describe the song from different aspects. Besides, the weights of affiliation matrix indicate the relatedness between users/items with taste clusters.

Second, to recommend track $i_{77414}$ to user $u_{71}$, ECF first finds their interactions of taste clusters according to the affiliations (\ie $c_1$, $c_4$), which are depicted as red circles in Figure~\ref{fig:casestudy-rec}. These taste clusters serve as a bridge to explicitly connect user's preference and items' semantics. Then two explanation paths ($i_{71} \rightarrow c_1 \rightarrow i_{77414}$ and $i_{71} \rightarrow c_4 \rightarrow i_{77414}$) are constructed for recommendation. For each path, we multiply their affiliation weight to generate the importance score of the taste cluster, according to Equation (\ref{equ:ecf-explanation-weight}). At last, by summarizing all the scores over intersected taste clusters, we can calculate the final prediction score, as well as the explanation paths from user $u_{71}$ to $i_{77414}$.

\subsubsection{\textbf{Taste Cluster Recommendation}}
Here we propose a new recommendation scenario, called \textit{taste cluster recommendation}, which means users are recommended with a bundle of similar items, and these items can be described with few tags (or features). We find this task is ubiquitous in real-world applications, like playlist recommendation in Spotify~\cite{recsys16spotify}. Current solutions to tackle this problem typically rely on manual selection by editors to guarantee the selected items are similar in tags (or other explicit features). However, this approach suffers from tedious selection process and poor personality. Our method could be a substitution for labor selections by learning taste clusters in an end-to-end manner. To be specific, since each learned taste cluster is a collection of items with descriptive tags, it can be naturally viewed as a playlist or game set. Thus, we can directly recommend taste clusters to users according to their affiliation weights. Taking user $u_{71}$ in Figure~\ref{fig:casestudy-rec} as an example, ECF could recommend him/her with a pop song playlist which is sung by female artists, since he/she used to listen lots of pop music by female artists, like Colbie Caillat, Adele and Celine Dion.

\subsubsection{\textbf{User Profiling}} 
The user-cluster affiliations discovered by ECF can also be used as user profiles directly. Take Figure~\ref{fig:casestudy-rec} for example, ECF assigns user $u_{71}$ to three clusters: $c_1$, $c_3$ and $c_4$. Then we can use the tag summary of these clusters as (part of) her profile. Such kind of clusters as user profiles can benefit various web applications, including (1) user-level predictive tasks, such as user demographic attributes prediction and user churn/uplift prediction \cite{huang2012customer,diemert2018large}, where cluster profiles can be consumed as auxiliary features; (2) ad audience targeting~\cite{cikm11ad,kdd09ad}, where ad platforms or advertisers can leverage user clusters to reach a certain segment of users that have targeted interests and habits in an interpretable manner; (3) look-alike audience extension~\cite{ma2016sub,liu2019real}, where the task is to reach users similar to a given set of seed users, especially with the goal of enhancing the performance of cold-start and long tail items and improving the recommendation diversity. User clusters are naturally suitable for this task because similar users have already been indexed in the same clusters.

%% file: 4_related.tex
\section{Related Works}

Collaborative filtering techniques, especially model-based CF methods, have achieved great success in personalized recommender systems. Most (if not all) of those methods rely on the \textit{latent} factor embeddings to generate rating predictions, which makes it difficult to give explanations for the predictions. However, explaining why a user likes an item can be as important as the accuracy of the prediction itself~\cite{20survey}, since it can not only enhance the transparency and trustworthiness of the system, but also significantly improve user satisfaction. Therefore, varieties of strategies for rendering explainable recommendations have been proposed. We first review explainable collaborative filtering  methods~\cite{www16emf,iui09tagsplanations,recsys13hft,sigir14efm,cikm15trirank,wsdm15flame,ijcai16rblt,www18atm,ijcai20amcf}, 
then we discuss other recent explainable recommendation works.




\textbf{Explainable collaborative filtering methods.} 
EMF~\cite{www16emf} adds an explainability regularizer into
the objective function of MF to force user/item hidden vectors to be close if a lot of the user’s neighbors also purchased the item. HFT~\cite{recsys13hft} leverages the topic model to obtain interpretable textual labels for latent rating dimensions. EFM~\cite{sigir14efm} factorizes a rating matrix in terms of both explicit phases from textual reviews as well as latent factors, and makes predictions by combing the scores of both explicit and implicit features. TriRank~\cite{cikm15trirank} models user-item-tag ternary relation as a heterogeneous tripartite graph and performs vertex ranking for recommendation. ATM~\cite{www18atm} utilizes latent factors to estimate tag importance of a user towards an item, and makes predictions via a linear combination of tag ratings. More recently, AMCF~\cite{ijcai20amcf} maps uninterpretable embeddings into the interpretable aspect features by minimizing both ranking loss and interpretation loss. Nonetheless, those methods are unable to satisfy all three important explanation properties: flexible, coherence and self-explainable, resulting in ineffectiveness to boost the transparency of recommender systems.

\textbf{Other directions of explainable recommendation.} There is also a large literature that considers other auxiliary information or other methods to explain recommendations~\cite{www16emf,www18narre,www18tem,sigir18mter,kdd18explanation,aaai19kg,sigir19cecf,ijcai19reconet,www21kgin,www22kgpath}. For example, NARRE~\cite{www18narre} designs an attention mechanism over the user and item reviews for rating prediction, and leverages the attention weight for explanation. 
VECF~\cite{sigir19cecf} proposes visually explainable recommendation based on personalized region-of-interest highlights by a multimodal attention network. PLM-Rec~\cite{www22kgpath} learns a language model over knowledge graph to generate explainable paths. Those methods leverage more expressive data for explanation, which can be the future direction of our ECF framework. 


%% file: 5_conclusion.tex
\section{Conclusion and future work}
In this paper, we propose a neat yet effective explainable collaborative filtering framework, called ECF, which is able to learn informative taste clusters in an end-to-end manner, and perform both recommendation and explanation by measuring the intersection between user-/item- cluster affiliations. ECF is flexible, coherent and self-explainable. Meanwhile, we design four quantitative metrics and conduct human studies to evaluate the explainability quality of taste clusters in a comprehensive manner. Extensive experiments conducted on three real-world datasets demonstrate the superiority of ECF from both recommendation accuracy and explanation quality. In the future, we aim to apply ECF to real-world recommendation scenarios with millions of users and items to improve its scalability. Besides, we plan to incorporate ECF with more expressive data beyond tags, such as reviews~\cite{www18narre}, knowledge graphs~\cite{www21kgin,HAKG22} to fully exploit the power of this framework.

%% file: 6_appendix.tex
\newpage
\appendix
\section{Appendix}


\renewcommand\thefigure{\Alph{section}\arabic{figure}}
\renewcommand\thetable{\Alph{section}\arabic{table}}    
\setcounter{figure}{0}
\setcounter{table}{0}

\begin{table*}[t]
    \caption{Complete Top-$K$ recommendation results. ``${\dagger}$'' indicates the improvement of the ECF over the baseline is significant at the level of 0.05. R and N refer to Recall and NDCG, respectively.}
    \centering
    \vspace{-5pt}
    \label{tab:appendix-top-k-recommendation}
    \resizebox{1.01\textwidth}{!}{
    \begin{tabular}{c|c c c c  |c c c c |c c c c }
    \hline
    \multicolumn{1}{c|}{\multirow{2}*{}}&
    \multicolumn{4}{c|}{Xbox} &
    \multicolumn{4}{c|}{MovieLen} &
    \multicolumn{4}{c}{Last-FM} \\
      &R@5 & R@10 &  N@5 & N@10 & R@5 & R@10 &  N@5 & N@10 & R@5 & R@10 &  N@5 & N@10 \\
    \hline
    MF  & 0.2615 & 0.3686 &  0.2383 & 0.2824  & 0.0601 & 0.0975  & 0.2738 & 0.2511 & 0.0289 & 0.0446  & 0.0428 & 0.0443\\
    NCF & 0.2372 & 0.3433  & 0.2065 & 0.2503  & 0.0594 & 0.0985  & 0.2701 & 0.2517  & 0.0269 & 0.0456  & 0.0396 & 0.0383  \\
    CDAE & 0.2604 & 0.3738  & 0.2346 & 0.2813  & 0.0609 & 0.0946  & 0.2671 & 0.2534  & 0.0286 & 0.0402  & 0.0431 & 0.0518  \\
    LightGCN & 0.2684 & 0.3625  & 0.2382 & 0.2837  & 0.0699 & 0.1163  & 0.2979 & 0.2752  & 0.0398 & 0.0578  & 0.0605 & 0.0634 \\ \hline
    EFM & 0.2647 & 0.3652  & 0.2368 & 0.2873   & 0.0657 & 0.1027 & 0.2866 & 0.2635  & 0.0319 & 0.0482  & 0.0471 & 0.0484 \\
    AMCF & 0.2601 & 0.3613  & 0.2355 & 0.2806  & 0.0603 & 0.0986  & 0.2719 & 0.2498  & 0.0295 & 0.0488  & 0.0456 & 0.0457 \\ \hline
    MF$_{forest}$ & \underline{0.2907} & \underline{0.3983} & \underline{0.2615} & \underline{0.3159} & \underline{0.0787} & \underline{0.1276} & \underline{0.3122} & \underline{0.2911} & \underline{0.0374} & \underline{0.0548} & \underline{0.0562} & \underline{0.0594} \\ \hline
    ECF$_{single}$ & 0.1714 & 0.2763 & 0.1423 & 0.1854  & 0.0352 & 0.0608  & 0.1584 & 0.1505  & 0.0205 & 0.0315  & 0.0339 & 0.0345 \\ 
    ECF & \textbf{0.2970}$^{\dagger}$ & \textbf{0.4299}$^{\dagger}$  & \textbf{0.2644}$^{\dagger}$ & \textbf{0.3193}$^{\dagger}$  & \textbf{0.0788} & \textbf{0.1325}$^{\dagger}$  & \textbf{0.3183}$^{\dagger}$ & \textbf{0.2952}$^{\dagger}$ & \textbf{0.0455}$^{\dagger}$ & \textbf{0.0635}$^{\dagger}$ & \textbf{0.0782}$^{\dagger}$ & \textbf{0.0749}$^{\dagger}$ \\
    \hline
    \end{tabular}}
    \vspace{-3pt}
\end{table*}

\subsection{Pseudocode of ECF}\label{appendix:pseudocode}
To help readers get an intuitive sense of the framework, we provide a pseudocode in Algorithm~\ref{alg:ecf}.

\begin{algorithm}[h]
\small
\caption{Training Pipeline for an individual ECF}
\label{alg:ecf}
\LinesNumbered
\KwIn{a set of users $\Set{U}$ and items $\Set{I}$; user-item interactions $\Mat{Y}$; tags $\Set{T}$ for items}
\KwOut{Learned user embeddings $\Mat{U}$; item embeddings $\Mat{V}$; taste cluster embeddings $\Mat{H}$; sparse affiliation matrix $\Mat{A}$ and $\Mat{X}$}

Randomly initialize all parameters \\
Randomly select $|\Set{C}|$ items as the initial taste clusters 

\While{not converge}{
\For{each pair $(u,i,j)$ in $\Mat{Y}$}{
    Calculate the sparse affiliations $\Mat{a}_u$, $\Mat{x}_i$ and $\Mat{x}_j$ with Eq.(\ref{equ:item-cluster-sparse})\\
    Perform cluster-based predictions with Eq. (\ref{equ:sparse-bpr}) \\
    Calculate tags distribution for taste clusters with Eq.(\ref{equ:tag-likelihood}) \\
    Calculate independence of taste clusters with Eq.(\ref{equ:mutual-information}) \\
    Update model parameters with Eq.(\ref{equ:ecf-loss})\\
}
}
\end{algorithm}

\subsection{Adaptation of EFM}\label{sec:appendix-efm}
Since EFM relies on the phase-level reviews for prediction and explanation, we make a few modifications to fit our settings where only tags of items are available. First, we generate the user-feature matrix by counting all the features of items that users have interacted with, which means that all the mentioned features are considered as positive. Second, the negative opinions of item's features are discarded because no auxiliary information is available. Third, the item-feature quality matrix is simplified as a 0-1 matrix where 1 denotes the item has the tag in attributes, while 0 denotes otherwise. After reconstructing the user-feature matrix and item-feature quality matrix, we run the EFM model as it describes in the paper.

\subsection{Hyper-Parameter Settings}\label{sec:appendix-hyerparameter}

We optimize all models with Adam~\cite{iclr14adam}, where the batch size is fixed at 1024. For CDAE, the hidden size is 100, and the dropout ratio is set to 0.5. For LightGCN, we set the number of layers to 3, and the dropout ratio is also set to 0.5. For EFM, we fix the percentage of explicit factors to 40\%, and tune the coefficient of feature-based score for best performance. For AMCF, we use their official codes (\url{https://github.com/pd90506/AMCF}) for implementation and set the hyperparameter of feature mapping as 0.05. For ECF, we set $m=20$ and $n=20$, which means that we restrict that each item and item are associated with only 20 clusters. The temperature hyperparameters $T$ and $\tau$ are all set as 2. We also set $|\Set{C}|=64$ for all datasets, which indicates each ECF model could generate 64 taste clusters. Besides, the number of single ECF model $F$ is set as 9 for all experiments. Moreover, early stopping strategy is performed for all methods, \ie premature stopping if recall@20 on the test set does not increase for 10 successive epochs. We report the average metrics for all the users in the testing set.

Besides, since ECF utilize sparse affiliations between users/items and taste clusters instead of embeddings to perform recommendation, we choose two options for fair comparison: (1) we fix the size of ID embeddings (users, items and taste clusters) as 64 for all methods; (2) since the affiliation size of ECF is 180 (top 20 affiliated taste clusters for each users/items, and we use 9 single ECF model), we align it with the embedding size of all other methods, and set the size of embeddings as 180. We conduct both experiments for all baselines and select the best performance for the report.

\subsection{Complete Recommendation Performance}\label{sec:appendix-topk}
We report the comprehensive recommendation performance in Table~\ref{tab:appendix-top-k-recommendation}, where $K=20$ is omitted since the results are already reported in Table~\ref{tab:top-k-recommendation}. From the results we see a similar trend as we discussed in Section~\ref{sec:exp-performance}: embedding-based methods (\eg MF and LightGCN) show strong performance across different datasets yet suffer from poor explanation; explainable methods can only slightly outperform MF, since they need to balance the trade-off between accuracy and transparency. However, ECF achieves the best performance in all ranking metrics (including the forest version of MF). Therefore, our model is able to not only provide accurate recommendation results, but also intuitive and clear explanations for each prediction with taste clusters.

\subsection{Baselines for Explanation Evaluation}\label{sec:appendix-baselines}

\begin{itemize}[leftmargin=*]
    \item \textbf{TagCluster} is a tag-oriented method which collects items according to their tags. To be specific, we first randomly sample $|\Set{C}|$ items as the seed of clusters, and search for the items which have the same tags with them. When there is no item that has the exactly the same tags as the seed, we relax it by randomly discarding one tag and continue to search. We iteratively conduct the operations until the size of clusters meets threshold $Z$. And we choose the 4 most frequent tags in the clusters as the tags of taste clusters.
    \item \textbf{K-means} is a similarity-oriented method which leverages the item embeddings from FM to perform K-means clustering algorithm. The result clusters are as directly used as the taste clusters. And we also select the top-4 most common tags as the descriptive tags for taste clusters.
    \item \textbf{Random} is a baseline method which randomly select $K$ items to construct the clusters. The procedure is repeated until all items are allocated in at least one cluster. We use the same mechanism as above methods to generate tags for each clusters.
\end{itemize}

\begin{table}[h]
    \caption{The human evaluation results on Last-FM dataset.}
    \centering
    \vspace{-10pt}
    \label{tab:app-human-lastfm}
    \resizebox{0.45\textwidth}{!}{
    \begin{tabular}{c|c|c|c|c}
    \hline
       &ECF & TagCluster & K-means & Random\\
    \hline
    RankOfTask(1) & \textbf{1.73} & 2 & 2.73 & 3.63 \\
    RankOfTask(2) & \textbf{1.3} & 2.5 & 2.23 & 3.93 \\
    \hline
    \end{tabular}}
    \vspace{-10pt}
\end{table}

\subsection{Human Evaluation}\label{sec:appendix-human}

We conducted two task to evaluate the explanability of ECF from human perspective. {\textbf{Task(1)} is to rank the quality of the clusters' tags generated by ECF and three baselines; and \textbf{Task(2)} is to rank the quality of user-item explanations, with rank 1 representing the best and rank 4 representing the worst. We randomly select 10 generated clusters for evaluation, and the mean ranking results of these two tasks are shown in Table~\ref{tab:app-human-lastfm}. We can see that ECF achieved the best from human judgements, which ranks 1.73 and 1.3 for clusters' tags and recommendation explanation, respectively. Since human evaluation can be time-consuming, expensive, and not scalable, our proposed computational metrics are important for automatic evaluation.

\subsection{Ablation Study \wrt $\lambda$}\label{sec:appendix-impact-lambda}
We also investigate the impact of different $\lambda$ on other two datasets, and the results are shown in Figure~\ref{fig:appendix-lambda}. We find that as the increase of auxiliary collaborative signals, the performance of ECF improves gradually. However, when the weighting coefficient $\lambda$ reaches beyond 0.6, the improvement becomes marginal. Thus, we set $\lambda=0.6$ for all experiments.

\subsection{Ablation Study \wrt Forest Mechanism}\label{sec:appendix-impact-forest}

We briefly discuss how the number of single ECF model influences the performance. Since the single ECF model is unable to model the complex and hidden users' interest space at once, more learned taste clusters may help to alleviate this issue. We confirm this assumption by adding more single ECF model for ensemble prediction, and the results on Xbox and MovieLens dataset are illustrated in Figure~\ref{fig:appendix-forest}.  We can see the huge performance gains when we increase the number of ECF models, especially from a single model to three models. However, as the number of single models continuously increases, the improvements become negligible. We set the number of ECF models to $9$ for all experiments.

\begin{figure}[h]
    \centering
    \vspace{10pt}
    \includegraphics[width=0.3\textwidth]{figures/lambda_legend.pdf}
    \vspace{3pt}
    \includegraphics[width=0.47\textwidth]{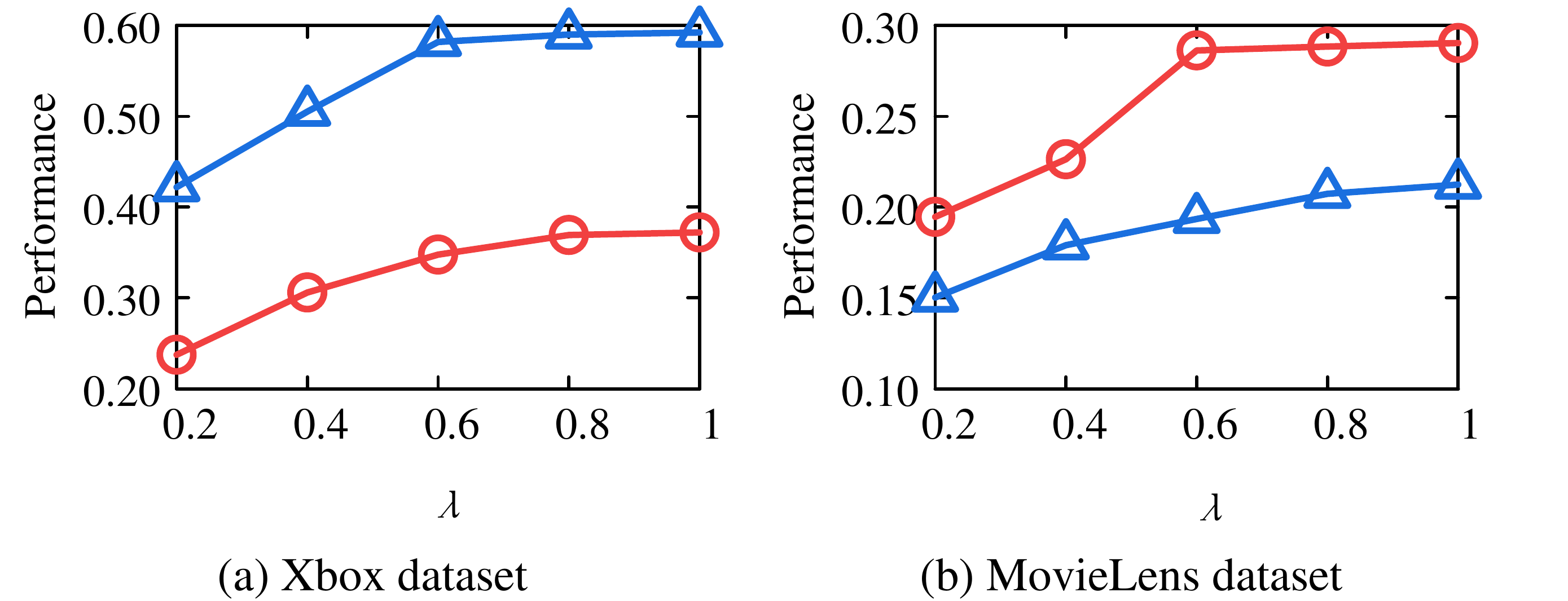}
    \vspace{-6pt}
    \caption{Impact of $\lambda$ on Xbox and MovieLens datsets.}
    \label{fig:appendix-lambda}
    \vspace{-5pt}
\end{figure}

\begin{figure}[h]
    \centering
    \vspace{5pt}
    \includegraphics[width=0.3\textwidth]{figures/lambda_legend.pdf}
    \vspace{3pt}
    \includegraphics[width=0.47\textwidth]{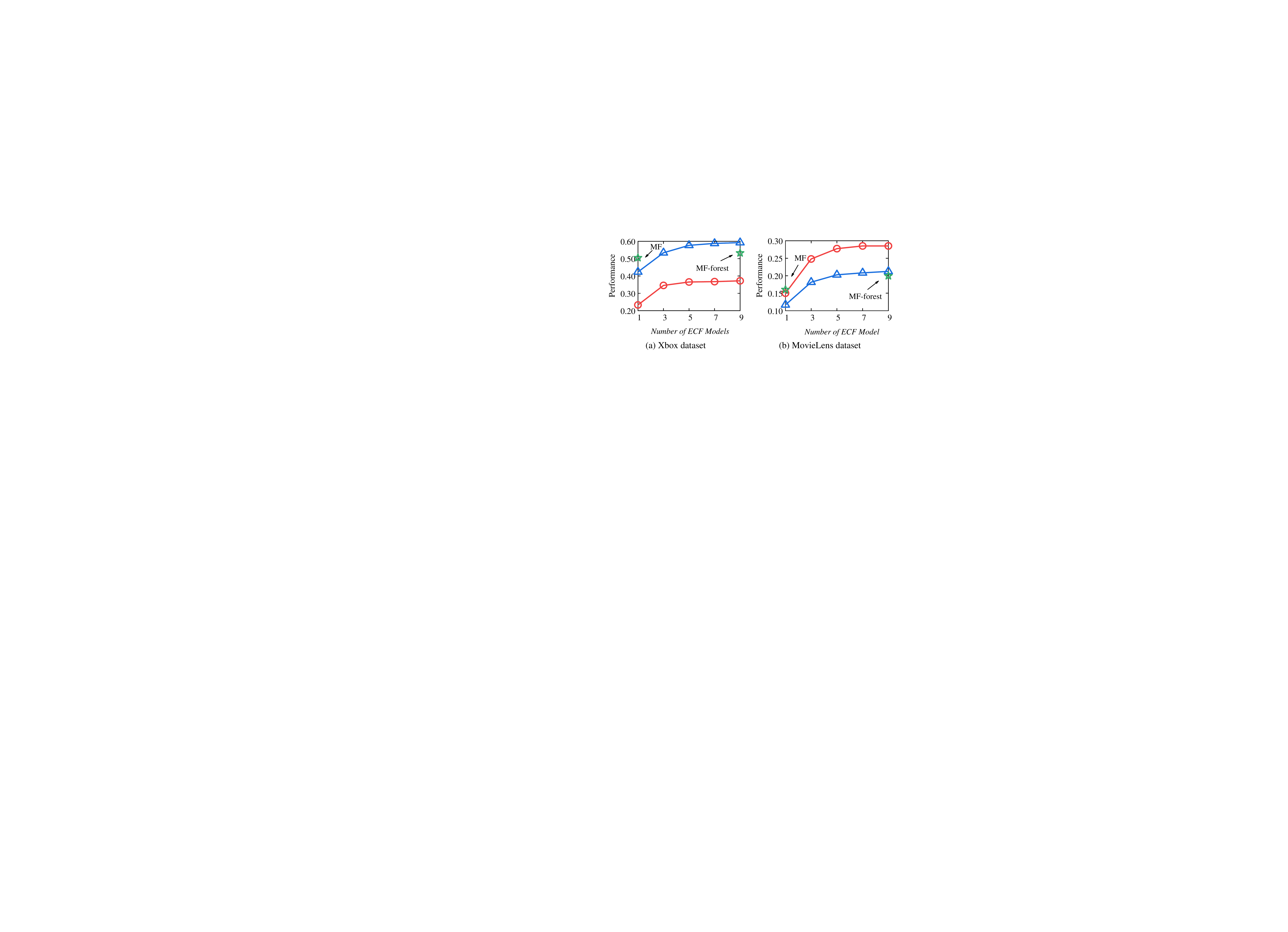}
    \vspace{-10pt}
    \caption{Impact of forest mechanism on Xbox and MovieLens datasets.}
    \label{fig:appendix-forest}
    \vspace{-3pt}
\end{figure}

\begin{table}[h]
    \caption{The recommendation performance and explainability of ECF and ECF$_{LGN}$ on Last-FM dataset.}
    \centering
    \vspace{-10pt}
    \label{tab:performance-ecf-lightGCN}
    \resizebox{0.48\textwidth}{!}{
    \begin{tabular}{c|c c |c c c c c}
    \hline
    \multicolumn{1}{c|}{\multirow{2}*{}}&
    \multicolumn{2}{c|}{Performance} &
    \multicolumn{5}{c}{Explainability} \\
       &R@20 & N@20 & Cov. & Util. & Sil. & Info. & Overall\\
    \hline
    ECF & 0.0851 & 0.0773 & 0.7648 & 0.6259 & 0.1584 & 0.2996 & 1.5352 \\
    ECF$_{LGN}$ & \textbf{0.0876} & \textbf{0.0792} & \textbf{0.7831} & \textbf{0.6430} & \textbf{0.1590} & \textbf{0.3042} & \textbf{1.5758} \\
    \hline
    \end{tabular}}
\end{table}

\subsection{Implementation of Discriminator}\label{sec:appendix-predictor}
We opt for a simple Multi-Layer Perceptron (MLPs) as the post-hoc discriminator. Specifically, it is a three-layer MLP (hidden size is 64) with ReLU as the activate function, cross-entropy as the loss function. We utilize the tags of items as the input data, and feed them into the MLPs to predict the item it belongs with. To improve the robustness of the model, we randomly mask 50\% of the item's tags for training. After training, we use the tags of taste clusters as the input, and collect top $|c_i|$ items as $R(\Set{T}_{c_i})$ by ranking their prediction probabilities.

\subsection{Implementation of ECF with LightGCN}\label{sec:appendix-ECF-LightGCN}
We use the item embeddings of the last layer in LightGCN for taste clusters learning, and the auxiliary collaborative signal is also replaced with the prediction loss from LightGCN. We denote this variant as ECF$_{LGN}$ and the results of both performance and explainability on Last-FM dataset are shown in Table~\ref{tab:performance-ecf-lightGCN}.
